\documentclass[journal]{IEEEtran}

\usepackage{xcolor,soul,framed} 

\colorlet{shadecolor}{yellow}
\usepackage[pdftex]{graphicx}
\graphicspath{{../pdf/}{../jpeg/}}
\DeclareGraphicsExtensions{.pdf,.jpeg,.png}

\usepackage[cmex10]{amsmath}
\usepackage{array}
\usepackage{mdwmath}
\usepackage{mdwtab}
\usepackage{eqparbox}
\usepackage{url}

\begin{document}
\bstctlcite{IEEEexample:BSTcontrol}
    \title{Plasmonic FET Terahertz Spectrometer}
  \author{Xueqing~Liu, Trond~Ytterdal,~\IEEEmembership{Senior Member,~IEEE},
      and~Michael~Shur,~\IEEEmembership{Life Fellow,~IEEE}

  \thanks{(Corresponding author: Michael Shur.)}
  \thanks{Xueqing Liu and Michael Shur are with the Department of Electrical, Computer and Systems Engineering, Rensselaer Polytechnic Institute, 110 8th Street, Troy, NY 12180, USA (e-mail: liux29@rpi.edu, shurm13@gmail.com).}
  \thanks{Trond Ytterdal is with the Department of Electronic Systems, Norwegian University of Science and Technology, 7491 Trondheim, Norway (e-mail: trond.ytterdal@ntnu.no).}
}


\maketitle

\begin{abstract}
We show that Si MOSFETs, AlGaN/GaN HEMTs, AlGaAs/InGaAs HEMTs, and p-diamond FETs with feature sizes ranging from 20\,nm to 130\,nm could operate at room temperature as THz spectrometers in the frequency range from 120\,GHz to 9.3\,THz with different subranges corresponding to the transistors with different features sizes and tunable by the gate bias. The spectrometer uses a symmetrical FET with interchangeable source and drain with the rectified THz voltage between the source and drain being proportional to the sine of the phase shift between the voltages induced by the THz signal between gate-to-drain and gate-to-source. This phase difference could be created by using different antennas for the source-to-gate and drain-to gate contacts or by using a delay line introducing a phase shift or even by manipulating the impinging angle of the two antennas. The spectrometers are simulated using the multi-segment unified charge control model implemented in SPICE and ADS and accounting for the electron inertia effect and the distributed channel resistances, capacitances and Drude inductances.
\end{abstract}


%
\IEEEpeerreviewmaketitle


\section{Introduction}

\IEEEPARstart{T}{erahertz} (THz) technology applications ranging from spectroscopy and imaging, non-destructive testing, quality control, and communications~\cite{Mittleman2018,Zeitler2017,Sun2017,Nellen2018,Zhong2018,Ahi2018,AfsahHejri2019,Shur201805,Shur201906,Rumyantsev2017} require sensitive detectors of THz and sub-THz radiation. Plasmonic field effect transistors (also called TeraFETs) have demonstrated excellent performance as THz and sub-THz detectors~\cite{Knap200212,Antonov200401,Teppe200508} and potential for THz generation~\cite{Knap200403,Yavorskiy201708}. A recent proposal is to use TeraFETs as spectrometers and interferometers of THz and sub-THz radiation based on the frequency-dependent THz signal rectification resulting from the phase difference in the THz voltages induced between the source-gate and drain-gate contacts of a single FET detector~\cite{Gorbenko2019}. The qualitative analytical theory presented in~\cite{Gorbenko2019} showed that the TeraFET spectrometer response varies from positive to negative with the gate bias. The gate voltage, at which the response is zero, depends on the frequency of the impinging THz signal that could be accurately determined. The strength of the response is proportional to $\sin{\theta}$, where $\theta$ is the phase difference between the signals coupled to the gate-to-source and gate-to-drain contacts. 

In this work, we simulate Si MOSFETs with feature sizes ranging from 20\,nm to 130\,nm and determine the TeraFET spectrometer operating ranges as functions of the device feature size. Our results show that such Si-based THz spectrometers could operate in the frequency range from 120\,GHz to 9.3\,THz. Using Si TeraFET that are fabricated using a standard Si VLSI technology opens up unique capabilities for cost-effective THz electronics technology that could enable Beyond 5G WiFi and THz communication systems.

\begin{figure}[!b]
  \begin{center}
  \includegraphics[width=\columnwidth]{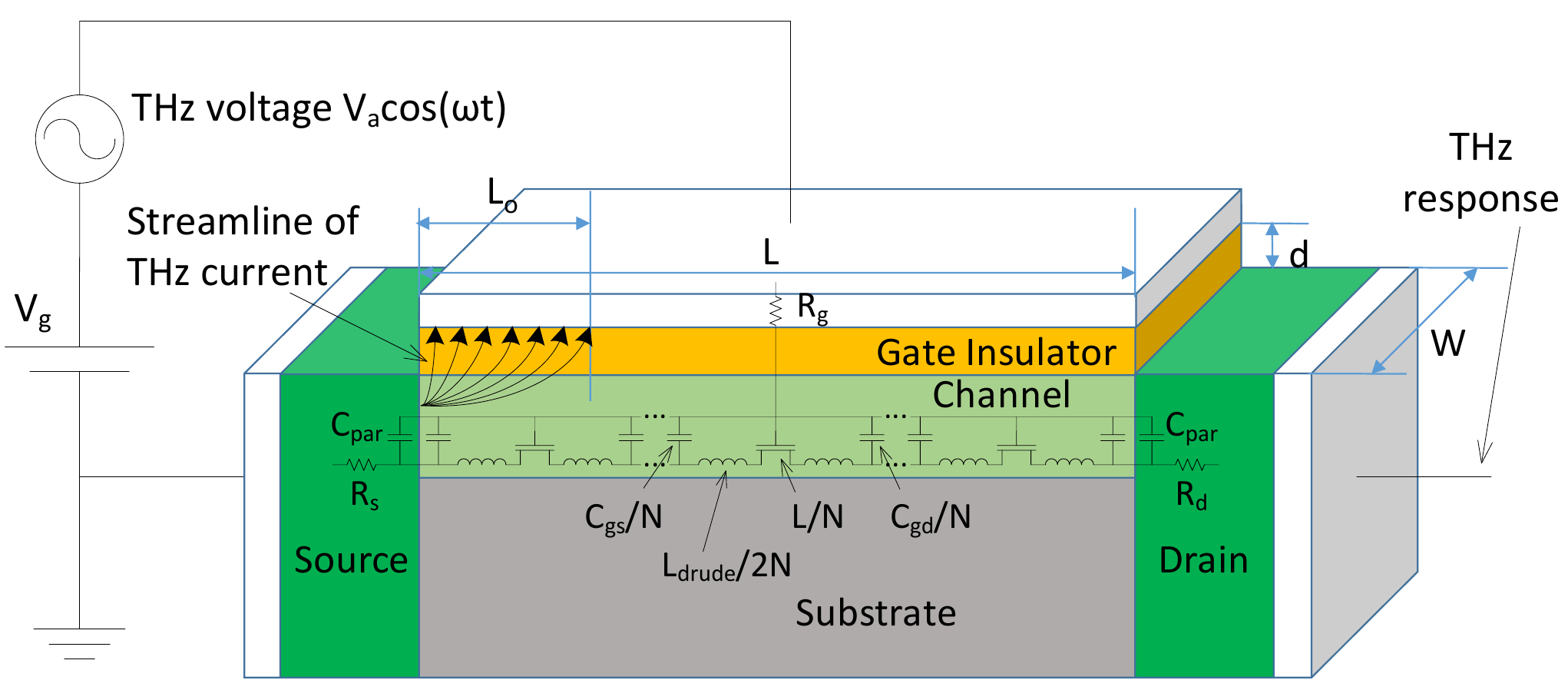}\\
  \caption{Plasmonic FET under THz radiation modeled with multiple segments in the channel accounting for the THz current distribution.}\label{fig1}
  \end{center}
\end{figure}

\begin{figure*}
  \begin{center}
  \includegraphics[width=\textwidth]{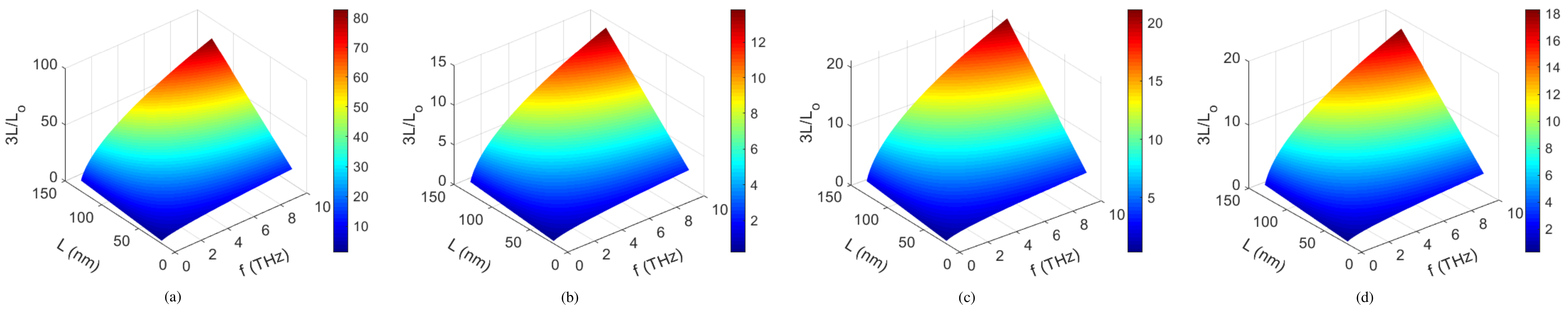}\\
  \caption{Minimum number of segments needed for the multi-segment model: (a) Si ($V_{gt}$= 0.11\,V, $\mu=$ 0.05\,m$^{2}$/Vs), (b) InGaAs ($V_{gt}$= 0.1\,V, $\mu=$ 0.35\,m$^{2}$/Vs), (c) GaN ($V_{gt}$= 0.1\,V, $\mu=$ 0.15\,m$^{2}$/Vs), (d) p-diamond ($V_{gt}$= 0.1\,V, $\mu=$ 0.2\,m$^{2}$/Vs).}\label{fig2}
  \end{center}
\end{figure*}

\begin{figure}
  \begin{center}
  \includegraphics[width=0.8\columnwidth]{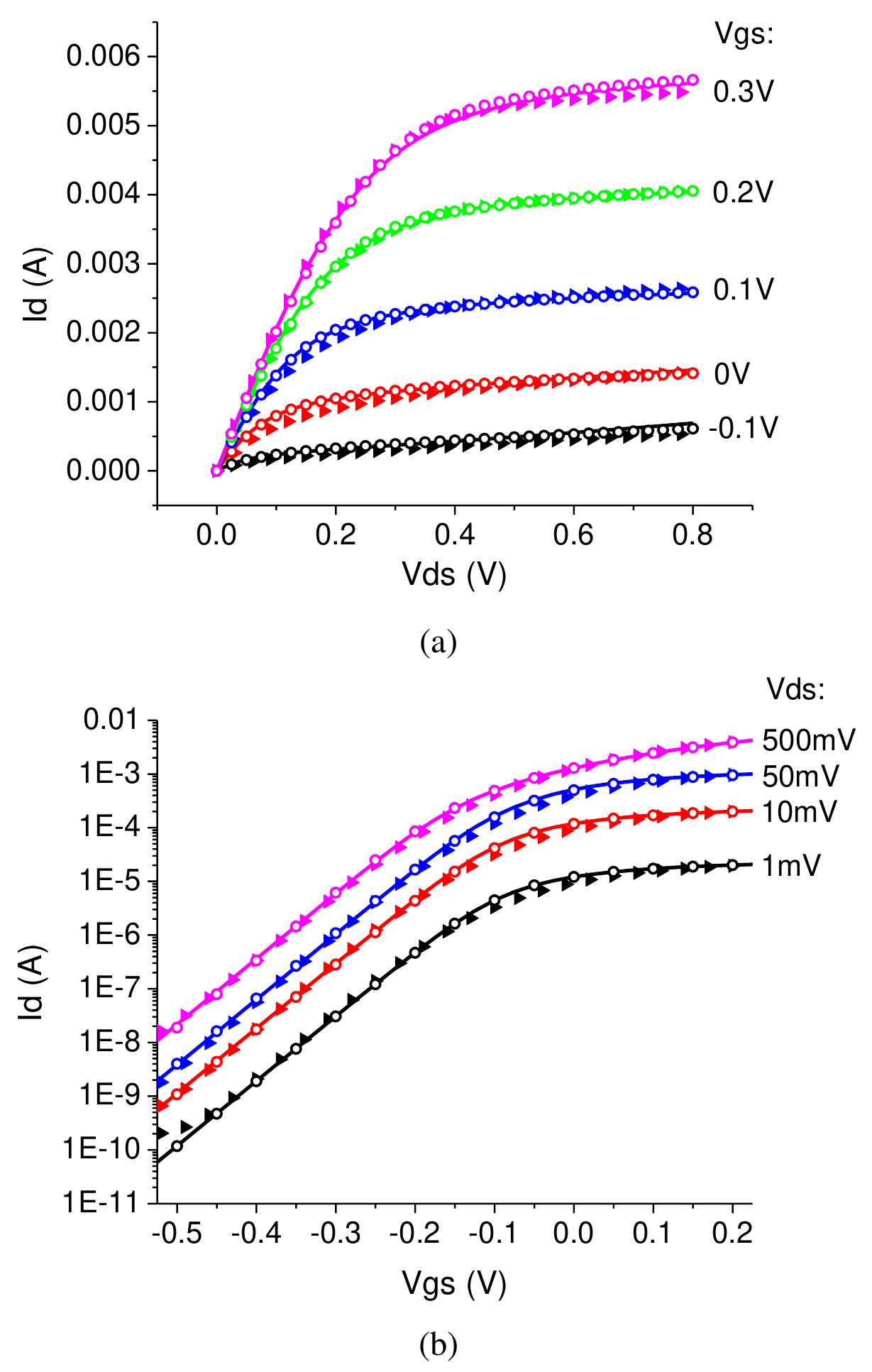}\\
  \caption{Comparison of the simulated I-Vs for the one segment SPICE model (lines) and the multi-segment model (circles) with the measured I-Vs (triangles) for AlGaAs/InGaAs pHEMTs~\cite{Liu201911}.}\label{fig3}
  \end{center}
\end{figure}

\begin{figure}
  \begin{center}
  \includegraphics[width=0.8\columnwidth]{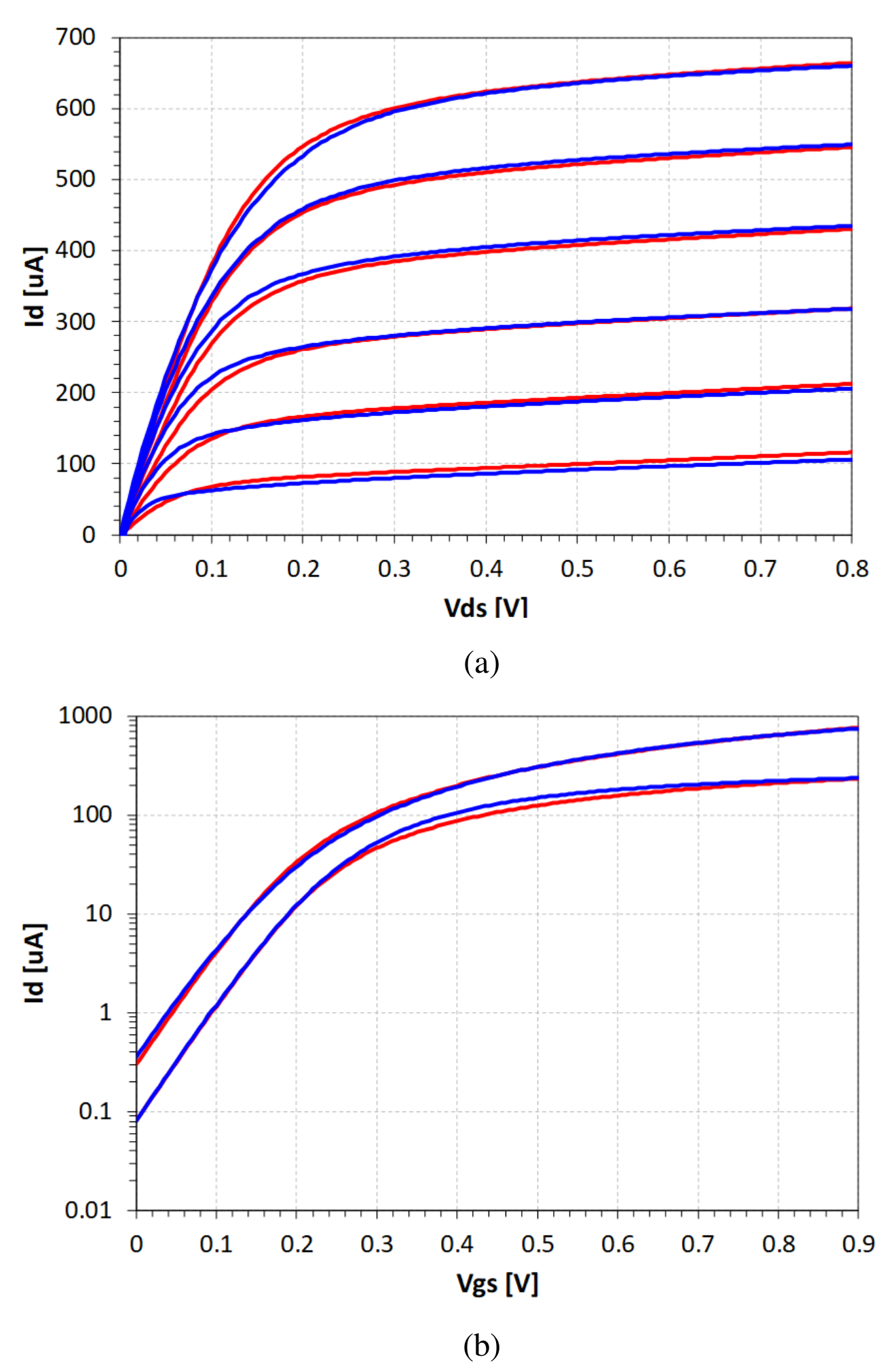}\\
  \caption{Comparison of the simulated I-Vs for the one segment SPICE model (red) with fitted to TCAD simulation results (blue) for the 20\,nm FDSOI NMOS ($W=$ 1\,$\mu$m and $L=$ 17\,nm): (a) output characteristics ($V_{gs}=$ 0.8\,V for the top curve and step is -0.1V) and (b) transfer characteristics ($V_{ds}=$ 0.6\,V for the top curve and $V_{ds}=$ 50\,mV for the bottom curve)~\cite{Liu201911}.}\label{fig4}
  \end{center}
\end{figure}

For the simulations, we use the unified charge control THz SPICE model for plasmonic field effect transistors implemented in Verilog-A~\cite{Liu201812,Lee1993}. It has been validated for FETs in various feature sizes and different material systems including 20\,nm FDSOI MOSFETs and 130\,nm AlGaAs/InGaAs pHEMTs~\cite{Liu201812,Lee1993}.

The THz spectrometer simulations using the SPICE model show that the signal detected as the drain-to-source voltage at the modulation frequency of the impinging THz radiation drops to zero at the frequency that is tunable by the gate voltage and establish the spectrometer frequency detection ranges that are functions of the gate length and the gate-source voltage.

\begin{figure*}
  \begin{center}
  \includegraphics[width=\textwidth]{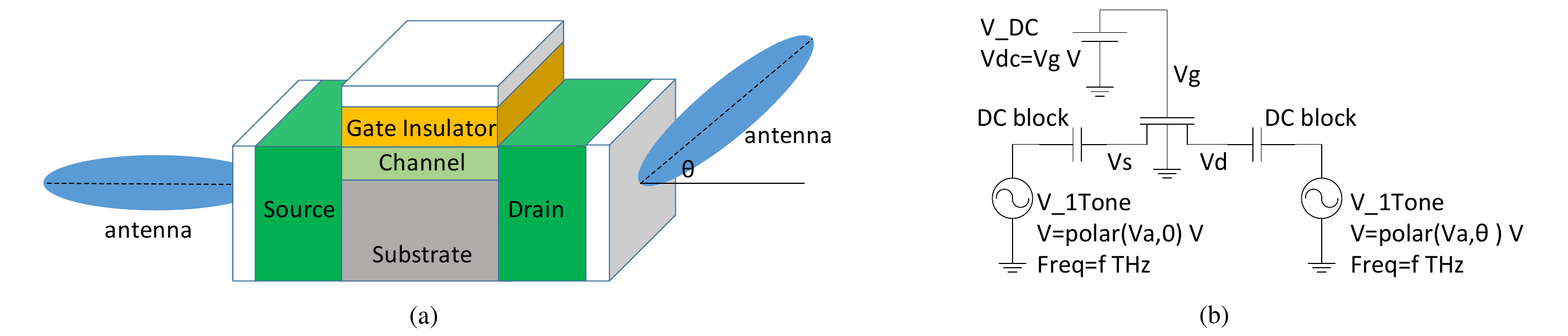}\\
  \caption{Schematics of operating a single plasmonic FET as a spectrometer~\cite{Gorbenko2019} (a) and the THz spectrometer simulation~\cite{Liu201911} (b).}\label{fig5}
  \end{center}
\end{figure*}

\begin{figure*}
  \begin{center}
  \includegraphics[width=\textwidth]{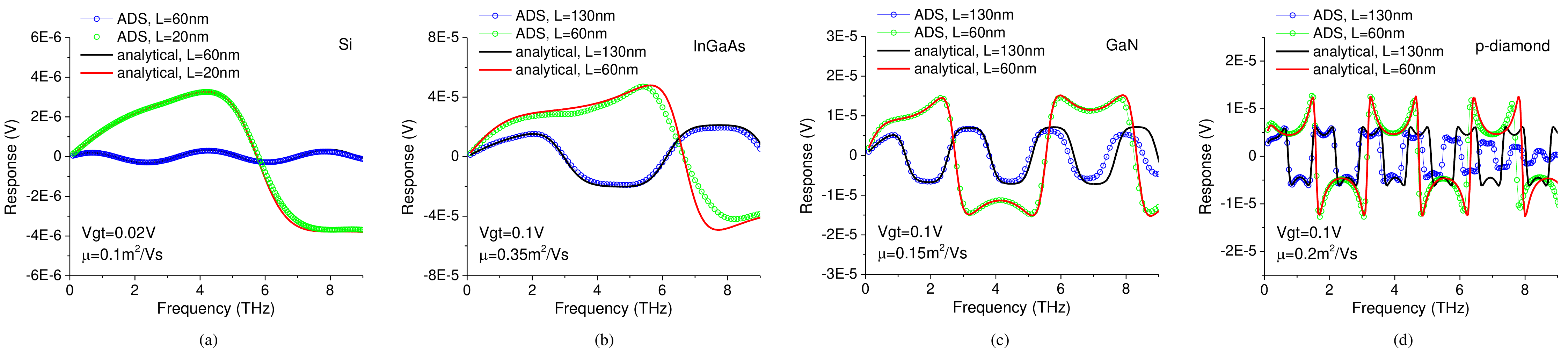}\\
  \caption{Spectrometer response as a function of frequency for (a) Si ($m=$ 0.19), (b) InGaAs ($m=$ 0.041), (c) GaN ($m=$ 0.24), and (d) p-diamond ($m=$ 0.74) without series resistance ($R_{g}=R_{s}=R_{d}=$ 0\,$\Omega$).}\label{fig6}
  \end{center}
\end{figure*}

\begin{figure*}
  \begin{center}
  \includegraphics[width=\textwidth]{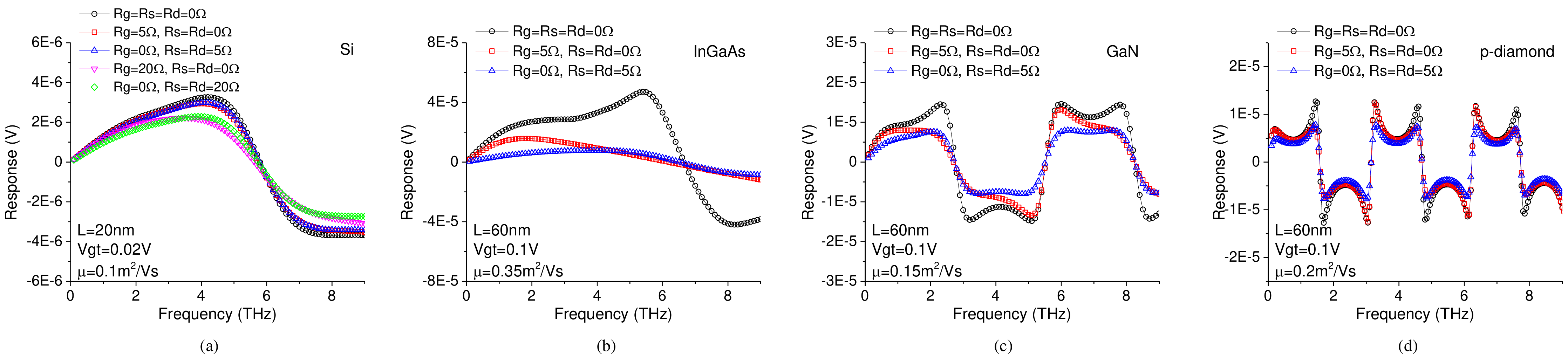}\\
  \caption{Effect of series resistance on the simulated spectrometer response as a function of frequency for (a) Si, (b) InGaAs, (c) GaN, and (d) p-diamond.}\label{fig7}
  \end{center}
\end{figure*}

\begin{figure*}[t]
  \begin{center}
  \includegraphics[width=\textwidth]{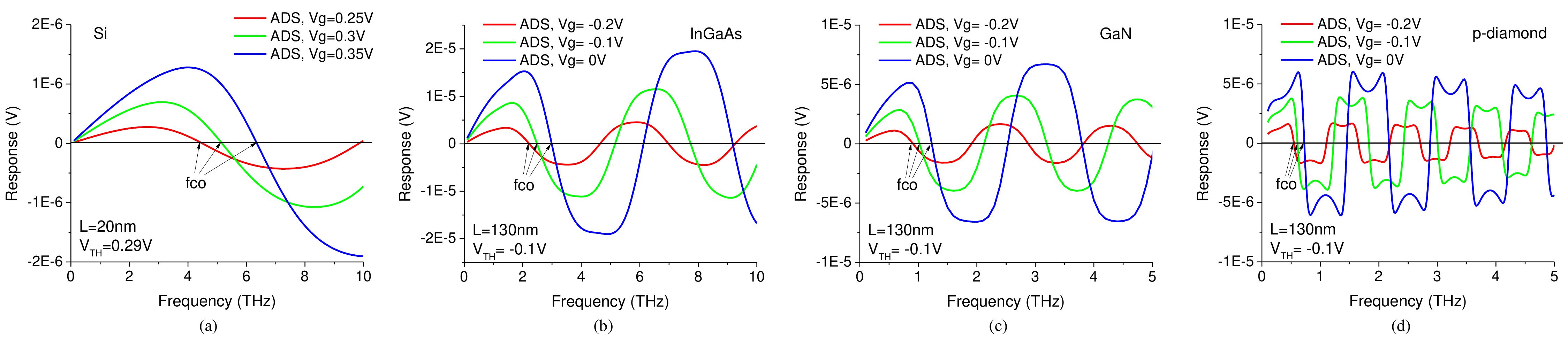}\\
  \caption{Simulated THz spectrometer response as a function of frequency at different gate biases for (a) Si ($\mu=$ 0.05\,m$^{2}$/Vs), (b) InGaAs ($\mu=$ 0.35\,m$^{2}$/Vs), (c) GaN ($\mu=$ 0.15\,m$^{2}$/Vs), (d) p-diamond ($\mu=$ 0.2\,m$^{2}$/Vs) without series resistance ($R_{g}=R_{s}=R_{d}=$ 0\,$\Omega$).}\label{fig8}
  \end{center}
\end{figure*}

\begin{figure*}
  \begin{center}
  \includegraphics[width=\textwidth]{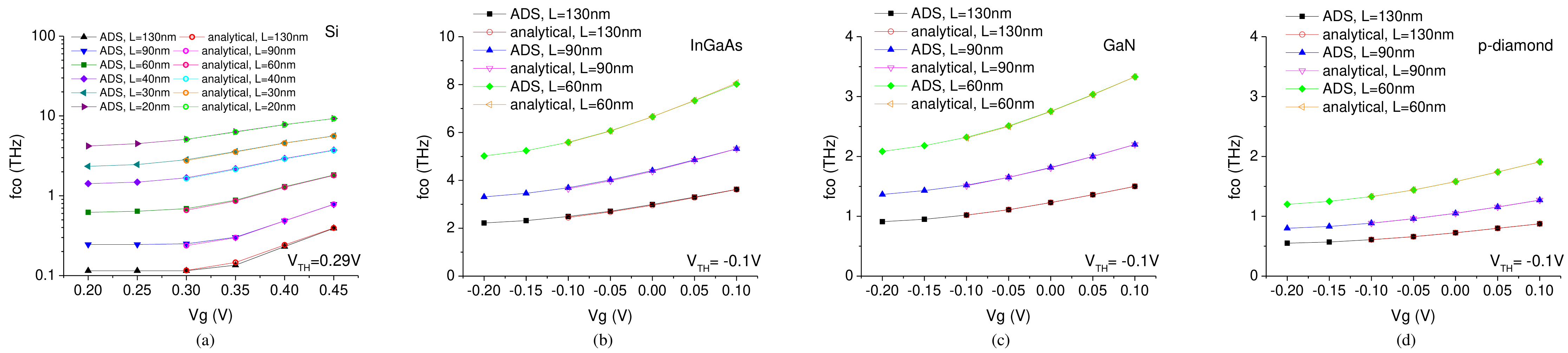}\\
  \caption{Extracted cross-over frequency as a function of gate bias for the SPICE model compared with the analytical results for (a) Si ($\mu=$ 0.05\,m$^{2}$/Vs), (b) InGaAs ($\mu=$ 0.35\,m$^{2}$/Vs), (c) GaN ($\mu=$ 0.15\,m$^{2}$/Vs), (d) p-diamond ($\mu=$ 0.2\,m$^{2}$/Vs) without series resistance ($R_{g}=R_{s}=R_{d}=$ 0\,$\Omega$).}\label{fig9}
  \end{center}
\end{figure*}

\begin{figure}
  \begin{center}
  \includegraphics[width=\columnwidth]{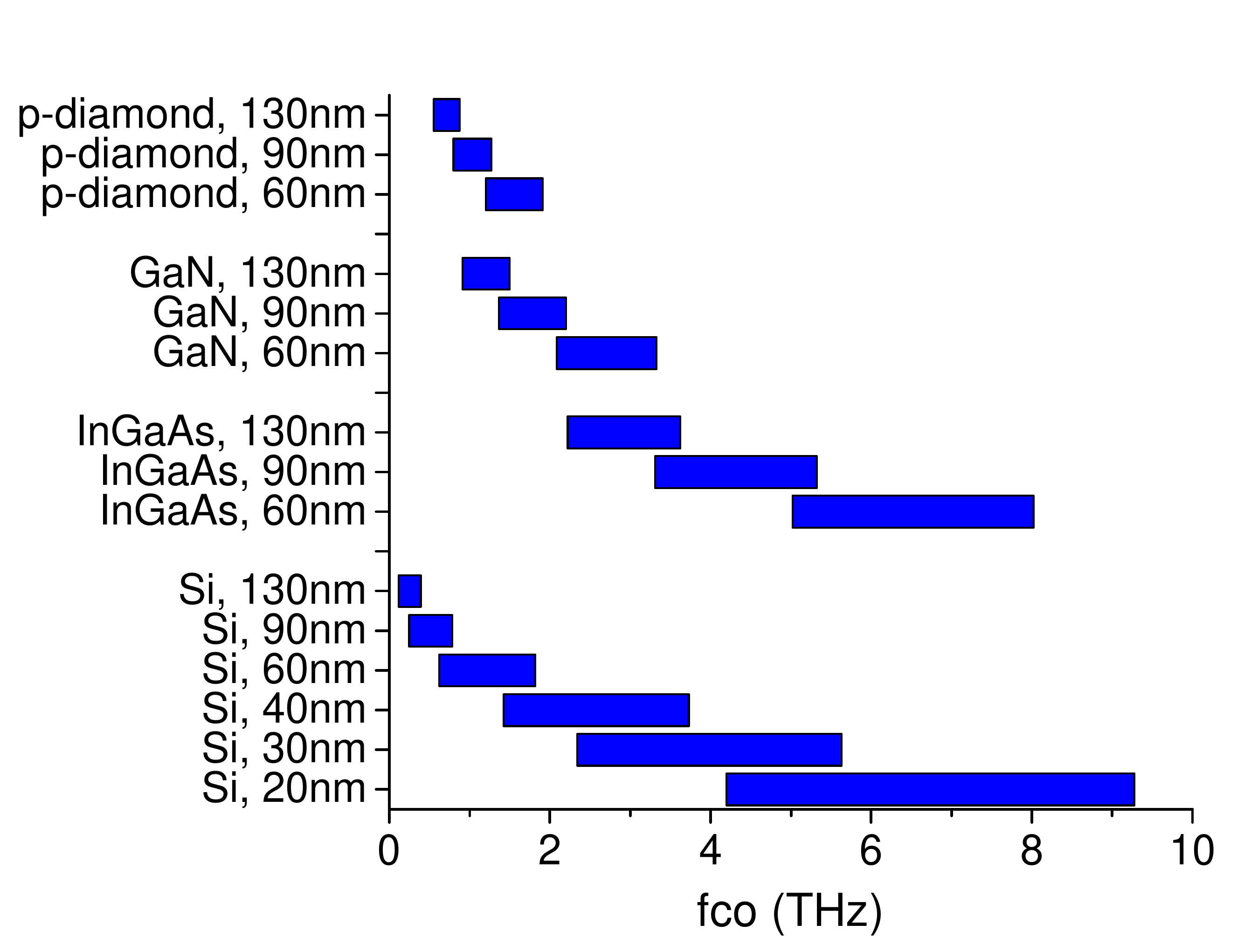}\\
  \caption{Comparison of the cross-over frequency range in different material systems with different feature sizes using the THz SPICE model.}\label{fig10}
  \end{center}
\end{figure}

\section{THz SPICE model}

The response of TeraFETs in the THz frequency range is the rectified drain-to-source voltage appearing due to the rectification of the decayed or resonant plasma waves in the FET channel~\cite{Dyakonov199310,Dyakonov199605,Dyakonov199610}. Fig.~\ref{fig1} illustrates the operating principle of a standard TeraFET THz detector with the transmission line in the device channel and the gate-to-channel current streamlines due to the impinging THz radiation. The impinging THz radiation couples to the FET via antennas connected to the gate-to-source and/or also to the gate-to-drain circuits or even just to the interconnects and contact pads. The excitations of the electron density – plasma waves – excited by the voltages at the THz frequencies due to the impinging radiation are rectified due to the nonlinear electron transport in the FET channel. The induced voltage across the FET channel is a DC drain-to-source voltage or (more practical) lower frequency voltage modulated due to the modulation of the impinging THz signal and measured by a lock-in amplifier to increase the signal-to-noise ratio. For $\omega \tau \ll 1$, where $\omega$ is the THz frequency and $\tau=m\mu/q$ is the momentum relaxation time, $\mu$ is the mobility, $m$ is the effective mass, and $q$ is the electronic charge, the plasma waves are overdamped. For $\omega \tau \gg 1$, the plasma waves are resonant~\cite{Knap200912}. For the standard TeraFET THz detector, the most efficient regime is when the THz signal is only coupled to the gate-to-source contacts as shown in Fig.~\ref{fig1}. The analytical THz detector theory was first derived for the above threshold regime~\cite{Dyakonov199605} and was then generalized to include the subthreshold regime and the parasitic resistances capacitances~\cite{Liu201812}.

Fig.~\ref{fig1} also shows the nonlinear transmission line representing the equivalent circuit in the transistor channel. In addition to the capacitances representing the gate-to-channel coupling and resistances accounting for the electron scattering in the device channel this improved equivalent circuit includes Drude inductances accounting for the electron inertia and important or even dominant at high frequencies~\cite{Liu201904,Liu201911}. Fig.~\ref{fig1} also schematically shows the THz current crowding near the gate edge due to low distributive capacitive impedance at high frequencies. 

The unified charge control model (UCCM) yields the equations for the intrinsic FET capacitances  $C_{gs}$ and $C_{gd}$~\cite{Lee1993}. The SPICE model accounts for the extrinsic components including the parasitic capacitances $C_{par}$ and the series resistances $R_{g}$, $R_{s}$ and $R_{d}$. The Drude inductance $L_{drude}=\tau R_{ch}$, where $R_{ch}$ is the channel resistance, accounts for the electron inertia~\cite{Shur201012} and, therefore, allows to describe the plasmonic resonances. To account for the THz current crowding, the channel is split into segments. The required  number of segments $N \ge 3L/L_{o}$, where $L$ is the channel length and $L_{o}=\sqrt{\mu V_{gt}/(2\pi f)}$, where $f$ is the THz radiation frequency, $V_{gt}=V_{gs}-V_{TH}$ is the gate voltage swing, $V_{TH}$ is the threshold voltage~\cite{Knap201403}. $L_{o}$ represents the characteristic scale of the voltage variation along the FET channel, which must be accounted at frequencies such that $L_{o} < L$ requiring the channel segmentation in the MOSFET model. Fig.~\ref{fig2} shows the minimum required number of the segments needed for different material systems as a function of the THz frequency and FET channel length.

Fig.~\ref{fig3} compares  the simulated I-V characteristic with the measured I-Vs for AlGaAs/InGaAs pHEMTs with 130\,nm gate length and 18\,$\mu$m gate width fabricated by Qorvo Inc. Fig.~\ref{fig4} compares the simulated I-V characteristics with the Sentaurus TCAD simulation results for a 20\,nm FDSOI NMOS ($W=$ 1\,$\mu$m and $L=$ 17\,nm). These comparisons validate our multi-segment UCCM SPICE/ADS model for different material systems~\cite{Liu201911}.

\section{Analytical model and SPICE simulation results for THz Spectrometer}

Fig.~\ref{fig5} (a) shows the schematic of the spectrometer using a single plasmonic FET~\cite{Gorbenko2019}. In the spectrometer regime of operation, the symmetry between the source and drain is broken by the phase shift $\theta$ of the THz voltages applied between the gate-source and gate-drain terminals.

In the above threshold regime, the rectified voltage across the MOSFET channel due to the impinging THz radiation is given by~\cite{Gorbenko2019}

\begin{equation}
V = \frac{\beta \omega V_{a}^2 \sin{\theta}}{4V_{gt} {\left\vert \sin {(kL)} \right\vert}^2 \sqrt{{\omega}^2 + {\gamma}^2}}.\label{eq1}
\end{equation}
Here $L$ is the channel length, $V_{a}$ is the THz voltage magnitude (the same between gate-source and gate-drain), $\omega=2\pi f$, $\gamma=1/\tau$, $k = (\Omega + i \Gamma) / s$ is the plasma wave vector, $\Omega = \sqrt{\sqrt{{\omega}^4 + {\omega}^2{\gamma}^2}/2 + {\omega}^2/2}$, $\Gamma = \sqrt{\sqrt{{\omega}^4 + {\omega}^2{\gamma}^2}/2 - {\omega}^2/2}$, $\beta = 8 \sinh{(\Gamma L/s)} \sin{(\Omega L/s)}$, $s$ is the plasma wave velocity.

Any one of the steady-state analysis types in the circuit simulator (SPICE or ADS) yields the spectrometer response for the THz spectrometer shown in Fig.~\ref{fig5} (b). Fig.~\ref{fig6} shows the calculated rectified drain-to-source voltage as a function of f for Si, AlGaAs/InGaAs, AlGaN/GaN and p-diamond FETs. Fig.~\ref{fig6} compares the analytical model and the ADS simulation for the 20-segment model. However, the analytical theory is not valid below threshold and does not account for parasitics that are important or even dominant for ultra-short channel FETs. Our compact multi-segment THz model resolves these issues and is suitable for the spectrometer design. Fig.~\ref{fig7} shows the effect of series resistance on the simulated spectrometer response. It could be seen that InGaAs based HFET is much more sensitive to the series resistance than other FETs especially Si MOS. 

Fig.~\ref{fig8} shows the effect of the gate bias on the THz spectrometer response. Fig.~\ref{fig9} shows the lowest cross-over frequency $f_{co}$ for different gate biases. Different FETs with different feature sizes could operate as the THz spectrometer in different frequency bands. For example, the subrange for the 20\,nm Si MOS spectrometer is from 4.2\,THz to 9.3\,THz. For Si MOS with feature sizes from 20\,nm to 130\,nm it is possible to cover the continuous THz band from 120\,GHz to 9.3\,THz, while the frequency bands covered for InGaAs, GaN and p-diamond FETs with feature sizes from 60\,nm to 130\,nm are from 2.2\,THz to 8\,THz, from 0.9\,THz to 3.3\,THz, and from 0.5\,THz to 1.9\,THz, respectively. Fig.~\ref{fig10} shows the comparison of the $f_{co}$ range in different material systems with different feature sizes using the THz SPICE model.

\section{Conclusion}

Using our compact multi-segment THz SPICE/ADS model accounting for the electron inertia and the distributed channel impedance, we simulated THz spectrometers using Si, InGaAs, GaN, and p-diamond FETs. Our results show that using the phase shift in the THz radiation coupling to the gate-to-source and gate-to-drain contacts, Si MOSFETs with feature sizes from 20\,nm to 130\,nm could operate as THz spectrometers in the 120\,GHz to 9.3\,THz frequency range, while InGaAs, GaN and p-diamond FETs with feature sizes from 60\,nm to 130\,nm could operate as the THz spectrometers in the 2.2\,THz to 8\,THz, 0.9\,THz to 3.3\,THz, and 0.5\,THz to 1.9\,THz frequency ranges, respectively. The spectrometers are tunable by the gate bias. This technology should enable novel THz components and systems for THz interferometry, imaging, and communications, including communications in Beyond 5G 240\,GHz to 300\,GHz range.

\section*{Acknowledgment}

The work at RPI was supported by the U.S. Army Research Laboratory Cooperative Research Agreement (Project Monitor Dr. Meredith Reed) and by the US AFOSR (Project Monitor Dr. Kenneth Goretta). The authors are grateful to Dr. V. Kachorovskii for useful discussions.


\begin{thebibliography}{00}

\bibitem{Mittleman2018} D. M. Mittleman, ``Twenty years of terahertz imaging,'' \emph{Opt. Express}, vol. 26, no. 8, pp. 9417--9431, Apr. 2018.

\bibitem{Zeitler2017} J. A. Zeitler, T. Rades, and P. F. Taday, ``Pharmaceutical and security applications of terahertz spectroscopy,'' in \emph{Terahertz Spectroscopy: Principles and Applications}, S. L. Dexheimer, Ed., 1\textsuperscript{st} ed. Boca Raton, FL, USA: CRC Press, 2017, pp. 323--348.

\bibitem{Sun2017} Q. Sun \textit{et al.}, ``Recent advances in terahertz technology for biomedical applications,'' \emph{Quant. Imaging. Med. Surg.}, vol. 7, no. 3, pp. 345--355, Jun. 2017.

\bibitem{Nellen2018} S. Nellen \textit{et al.}, ``Recent progress of continuous-wave terahertz systems for spectroscopy, non-destructive testing, and telecommunication,'' in \emph{Proc. SPIE, Terahertz, RF, Millimeter, and Submillimeter-Wave Technology and Applications XI}, Vol. 10531, Feb. 2018, Art. no. 105310C.

\bibitem{Zhong2018} S. Zhong, ``Progress in terahertz nondestructive testing: A review,'' \emph{Front. Mech. Eng.}, vol. 14, no. 3, pp. 273--281, May 2018.

\bibitem{Ahi2018} K. Ahi, S. Shahbazmohamadi, and N. Asadizanjani, ``Quality control and authentication of packaged integrated circuits using enhanced-spatial-resolution terahertz time-domain spectroscopy and imaging,'' \emph{Opt. Lasers Eng.}, vol. 104, pp. 274--284, May 2018.

\bibitem{AfsahHejri2019} L. Afsah‐Hejri \textit{et al.}, ``A Comprehensive Review on Food Applications of Terahertz Spectroscopy and Imaging,'' \emph{Compr. Rev. Food Sci. Food Saf.}, vol. 18, no. 5, pp. 1563--1621, Sep. 2019.

\bibitem{Shur201805} M. Shur, ``Plasmonic Detectors and Sources for THz Communication and Sensing,'' in \emph{Proc. SPIE, Micro- and Nanotechnology Sensors, Systems, and Applications X}, Vol. 10639, May 2018, Art. no. 1063929.

\bibitem{Shur201906} M. Shur \textit{et al.}, ``TeraFETs for Terahertz Communications,'' \emph{Photonics Newsletter}, vol. 33, no. 3, pp. 4--7, Jun. 2019.

\bibitem{Rumyantsev2017} S. Rumyantsev \textit{et al.}, ``Terahertz Beam Testing of Millimeter Wave Monolithic Integrated Circuits,'' \emph{IEEE Sensors J.}, vol. 17, no. 17, pp. 5487--5491, Jul. 2017.

\bibitem{Knap200212} W. Knap, Y. Deng, S. Rumyantsev, and M. S. Shur, ``Resonant detection of subterahertz and terahertz radiation by plasma waves in submicron field-effect transistors,'' \emph{Appl. Phys. Lett.}, vol. 81, no. 24, pp. 4637–-4639, Dec. 2002.

\bibitem{Antonov200401} A. V. Antonov \textit{et al.}, ``Electron transport and terahertz radiation detection in submicrometer-sized GaAs/AlGaAs field-effect transistors with two-dimensional electron gas,'' \emph{Phys. Solid State}, vol. 46, no. 1, pp. 146--149, Jan. 2004

\bibitem{Teppe200508} F. Teppe \textit{et al.}, ``Room-temperature plasma waves resonant detection of subterahertz radiation by nanometer field-effect transistor,'' \emph{Appl. Phys. Lett.}, vol. 87, no. 5, Aug. 2005, Art. no. 052107.

\bibitem{Knap200403} W. Knap \textit{et al.}, ``Terahertz emission by plasma waves in 60 nm gate high electron mobility transistors,'' \emph{Appl. Phys. Lett.}, vol. 84, no. 13, pp. 2331--2333, Mar. 2004.

\bibitem{Yavorskiy201708} D. Yavorskiy \textit{et al.}, ``Sub-terahertz emission from field-effect transistors,'' \emph{Acta Physica Polonica A}, vol. 132, no. 2, pp. 335--337, Aug. 2017.

\bibitem{Gorbenko2019} I. V. Gorbenko, V. Y. Kachorovskii, and M. Shur, ``Terahertz plasmonic detector controlled by phase asymmetry,'' \emph{Opt. Express}, vol. 27, no. 4, pp. 4004--4013, Feb. 2019.

\bibitem{Liu201812} X. Liu, K. Dovidenko, J. Park, T. Ytterdal, and M. S. Shur, ``Compact Terahertz SPICE Model: Effects of Drude Inductance and Leakage,'' \emph{IEEE Trans. Electron Devices}, vol. 65, no. 12, pp. 5350--5356, Dec. 2018.

\bibitem{Lee1993} K. Lee, M. Shur, T. Fjeldly, and T. Ytterdal, \emph{Semiconductor device modeling for VLSI}, Princeton, NJ, USA: Prentice Hall, 1993, pp. 441--478.

\bibitem{Dyakonov199310} M. Dyakonov and M. S. Shur, ``Shallow water analogy for a ballistic field effect transistor: New mechanism of plasma wave generation by dc current,'' \emph{Phys. Rev. Lett.}, vol. 71, no. 15, pp. 2465--2468, Oct. 1993.

\bibitem{Dyakonov199605} M. I. Dyakonov and M. S. Shur, ``Detection, mixing, and frequency multiplication of terahertz radiation by two-dimensional electronic fluid,'' \emph{IEEE Trans. Electron Devices}, vol. 43, no. 3, pp. 380--387, Mar. 1996. 

\bibitem{Dyakonov199610} M. I. Dyakonov and M. S. Shur, ``Plasma wave electronics: novel terahertz devices using two-dimensional electron fluid,'' \emph{IEEE Trans. Electron Devices}, vol. 43, no. 10, pp. 1640--1645, Oct. 1996.

\bibitem{Knap200912} W. Knap \textit{et al.}, ``Field effect transistors for terahertz detection: Physics and first imaging applications,'' \emph{J. Infrared Millim. Terahertz Waves}, vol. 30, no. 12, pp. 1319--1337, Dec. 2009.

\bibitem{Liu201904} X. Liu, T. Ytterdal, V. Yu. Kachorovskii, and M. S. Shur, ``Compact terahertz SPICE/ADS model,'' \emph{IEEE Trans. Electron Devices}, vol. 66, no. 6, pp. 2496--2501, Apr. 2019.

\bibitem{Liu201911} X. Liu, T. Ytterdal, and M. Shur, ``Silicon MOSFET Terahertz Spectrometer,'' submitted to \emph{Int. Microw. Symp. (IMS)}.

\bibitem{Shur201012} M. Shur, ``Plasma wave terahertz electronics,'' \emph{Electron. Lett.}, vol. 46, no. 26, pp. 18--21, Dec. 2010.

\bibitem{Knap201403} W. Knap \textit{et al.}, ``Recent Results on Broadband Nanotransistor Based THz Detectors,'' in \emph{Proc. THz. Secur. Appl.}, Mar. 2014, pp. 189--209.

\end{thebibliography}
\end{document}